\documentclass[a4paper]{revtex4}
\usepackage{graphicx}
\usepackage{fancyhdr}
\usepackage{amsmath}
\usepackage{color}

\begin{document}

\title{The classically conformal B$-$L extended Ma model}


\author{Y. Orikasa}
\affiliation{School of Physics, KIAS, Seoul 130-722, Korea}
\affiliation{Department of Physics and Astronomy, Seoul National University, Seoul 151-742, Korea}

\begin{abstract}
We discuss a classically conformal radiative seesaw model with gauged B$-$L symmetry, in which the 
B$-$L symmetry breaking can occur through the Coleman-Weinberg mechanism.
 As a result, all mass terms are generated and
EW symmetry breaking also occurs. 
We show some allowed parameters to satisfy several theoretical constraints. 
\end{abstract}

\maketitle

\thispagestyle{fancy}

\section{Introduction}
The standard model (SM) has to be still extended so as to include massive neutrinos  and dark matter (DM), 
even though the SM Higgs has been discovered. 
One of the elegant solutions to resolve this issue is known as radiative seesaw models
~\cite{Zee,Li, Ma:2006km, Kanemura:2011vm, Kanemura:2011mw, Hehn:2012kz}, 
in which active neutrino masses are generated 
at multi-loop level and exotic fields are naturally introduced in order to get such radiative masses. 
Such an exotic field can be often identified as a DM candidate. 
Especially, Ma model~\cite{Ma:2006km} is known as a minimal radiative seesaw model including fermionic and bosonic DM candidates.

%%%%%%%%%  Conformal Model %%%%%%%%%%%%%  
On the other hand the hierarchy problem arises in the SM. 
One solution of the hierarchy problem is supersymmetry. 
This is a beautiful theory however there are no signals in the LHC experiments. 
%In these days, the alternative solutions are discussed \cite{Aoki:2012xs}.
In this paper, we take another approach to the hierarchy problem following Bardeen's argument\cite{Bardeen:1995kv}.
Bardeen has argued that once the classically conformal symmetry and its minimal violation 
by quantum anomalies are imposed on the SM, it may be free from quadratic divergences. 
The models based on this idea are called classically conformal models\cite{Hempfling:1996ht, Meissner:2006zh, 
Iso:2009ss, Iso:2009nw, Iso:2012jn, Hashimoto:2013hta, Hashimoto:2014ela}. 
The classical Lagrangian for these models has no mass terms and all dimensional parameters are dynamically generated. 
The models need an absence of intermediate scales between the TeV scale and Planck scale. 
Then the Planck scale physics is directly connected to the electroweak (EW) physics.  

We consider the classically conformal Ma model which is combined Ma model and the classically conformal model. 
This model connects tiny neutrino mass scale and Planck scale. 
However the minimal classically conformal Ma model doesn't realize for following two reasons. 
First the EW symmetry does't occur for large top Yukawa coupling. 
Second the classically conformal symmetry forbids Majorana mass term. 
The Majorana mass plays an important role in Ma model. 
We need the extended model. 
The minimal extension is gauged B$-$L model. 
In this model, EW symmetry breaking is triggered by B$-$L symmetry breaking and 
Majorana mass term is generated by B$-$L symmetry breaking. 

This paper is based on our work\cite{Okada:2014nea}.

%%%%%%%%%%%%%%%%%%%%%%%%%%%%%%%%%%
\section{The classically conformal B$-$L extended Ma model}

\begin{table}[thbp]
\centering {\fontsize{10}{12}
\begin{tabular}{|c||c|c|c|}
\hline Fermion & $L_L$ & $ e_{R} $ & $N_R$  
  \\
  \hline
$(SU(2)_L,U(1)_Y)$ & $(\bm{2},-1/2)$ & $(\bm{1},-1)$ & $(\bm{1},0)$  
\\\hline
%$U(1)_L$ & $+1$ & $-1$  & $-1$ & $+1$ & $-1$ & $0$ & $0$  & $0$  \\\hline
%%%
%$U(1)'$ & $-1/2$ & $-1/2$ & $-1/2$ & $-1/2$ & $3/2$  \\\hline
%$U(1)'$ & $\ell$ & $\ell$ & $\ell$ & $\ell$ & $\ell-\eta_2$ & $\ell-\eta_2$  \\\hline
$U(1)_{B-L} $ & $-1$ & $-1$ &  $-1$    \\\hline
$Z_2$ & $+$ & $+$ &  $-$  \\\hline
%%%
\end{tabular}%
} \caption{$L_L$, $e_R$, and $N_R$ have three generations, which is abbreviated.} 
\label{tab:1}
\end{table}

\begin{table}[thbp]
\centering {\fontsize{10}{12}
\begin{tabular}{|c||c|c|c|}
\hline Boson  & $\Phi$   & $\eta$    & $\varphi$ 
  \\
  \hline
$(SU(2)_L,U(1)_Y)$ & $(\bm{2},1/2)$  & $(\bm{2},1/2)$   & $(\bm{1},0)$ \\\hline
%$U(1)_L$ & $0$ & $0$ & $0$  & $0$  \\\hline
%%%
%$U(1)' $  & $0$ & $0$ & $1$  & $-1$   & $2$  \\\hline
%$U(1)' $  & $0$ & $0$ & $\eta_2$  & $\eta_2-2\ell$   & $-2\ell$  \\\hline
$U(1)_{B-L} $ & $0$ & $0$ &  $2$    \\\hline
$Z_2$ & $+$ & $-$ &  $+$  \\\hline
%%%
\end{tabular}%
} 
\caption{The particle contents  for bosons. }
\label{tab:2}
\end{table}
%%%%%%%%%%%%%%%%%%%%%%%%%%%%%%%%%%

We discuss the one-loop induced radiative neutrino model with gauged $U(1)_{B-L}$ symmetry 
containing the DM candidates: the lightest field of $N_R$ and $\eta$ which $Z_2$ odd are assigned. 
The particle contents are shown in Tab.~\ref{tab:1} and Tab.~\ref{tab:2}. 
We add three $SU(2)_L$ singlet Majorana fermions $N_R$ with $-1$ charge under the $B-L$ symmetry to the SM fields.
For new bosons, we introduce a $SU(2)_L$ doublet scalar $\eta$  with zero charge under the $B-L$ symmetry, and a neutral $SU(2)_L$ singlet scalar $\varphi$ with  $2$ charge under the $B-L$ symmetry to the SM fields.
%%%
We assume that  the SM-like Higgs $\Phi$ and $\varphi$ have respectively  vacuum
expectation value (VEV); $v/\sqrt2$ and $v'/\sqrt2$.

The relevant  Lagrangian for Yukawa sector and scalar potential under these assignments
are given by
\begin{eqnarray}
-\mathcal{L}_{Y}
&=&
(y_\ell)_a \bar L_{La} \Phi e_{Ra} + (y_{\eta})_a \bar L_{La} \eta^*   N_{Ra}
+\frac12y_{N} \varphi \bar N^c_R N_R 
+\rm{h.c.} \label{Lag:Yukawa}\\ 
%%%
\mathcal{V}
&=& 
  \lambda_\Phi |\Phi|^4 
  %%%
  + \lambda_\eta |\eta|^4 
  + \lambda_\varphi |\varphi|^4
 %%%
  + \lambda_{\Phi\eta} |\Phi|^2 |\eta|^2
  + \lambda'_{\Phi\eta}  |\Phi^\dagger \eta|^2
  + \lambda''_{\Phi\eta}  [(\Phi^\dag\eta)^2+{\rm c.c.}]\nonumber\\
  &&
  + \lambda_{\Phi\varphi}  |\Phi|^2 |\varphi|^2 + \lambda_{\eta\varphi} | \eta|^2 |\varphi|^2
,
\label{HP}
\end{eqnarray}
where mass terms are forbidden by the conformal symmetry, $a=1 \mathchar`-3$, and the first term of $\mathcal{L}_{Y}$ can generates the (diagonalized) charged-lepton masses. 
Without loss of generality, we here work on the basis that the third term of $\mathcal{L}_{Y}$ is diagonalized and of $y_{N}$ is real and positive.

\subsection{Symmetry breakings}
We discuss the symmetry breakings in our model. 
We assume the classically conformal symmetry and the EW symmetry breaking doesn't occur by 
negative mass parameter.
The symmetry breaking is occurred by radiatively\cite{Coleman:1973jx}. 

We assume the following conditions at the Planck scale for simplicity, 
\begin{eqnarray}
\lambda_{\Phi\eta}=\lambda_{\Phi\eta}^\prime=\lambda_{\Phi\varphi}=\lambda_{\eta\varphi}=0. 
\end{eqnarray}
Under this assumption, these couplings are generated by quantum correction. As a result,  the couplings 
are very small at low energy scale.  Therefore we can consider the SM with inert doublet sector and the B$-$L 
sector separately. 

%First, we consider the B$-$L sector. 
%The B$-$L symmetry is broken by the Coleman-Weinberg mechanism\cite{Coleman:1973jx}. In this scenario, the running coupling 
%$\lambda_{\varphi}$ should satisfy the following relation at the symmetry breaking scale, 
The Coleman-Weinberg condition is the following, 
\begin{align}
\lambda_{\varphi}(\mu=v') \sim -\frac{3}{4\pi^2}\left(g^4_{B-L} 
 -\frac{1}{96}Tr\left[y_N^\dagger y_N y_N^\dagger y_N\right] \right).
 \label{CWcon}
\end{align}
These running couplings should be satisfy this condition at the symmetry breaking scale($v'$). 
The $\Phi$ mass can be obtained by the following form, 
\begin{eqnarray}
 m_\varphi^2=-4\lambda_\varphi v'^2. 
\end{eqnarray}
This mass should be positive for the stable vacuum. 
It suggests $\lambda_\varphi$ is negative at the B$-$L symmetry breaking scale.  

Once the B$-$L symmetry is broken, all mass terms are generated through the B$-$L breaking scale. 
The SM Higgs doublet mass is generated through the mixing term 
between the SM Higgs and B$-$L breaking scalar in the potential. 
The effective tree-level mass squared is induced. If $\lambda_{\Phi\varphi}$ is negative, 
the EW symmetry breaking occurs as usual in the SM. 
Under our assumption($\lambda_{\Phi\varphi}(M_{pl})=0$), $\lambda_{\Phi\varphi}$ becomes negative 
at the B$-$L breaking scale because of positive RGE. 
Inserting the tadpole condition, $\lambda_\Phi=-\lambda_{\Phi\varphi} v'^2/(2v^2)$, the SM Higgs mass 
is given by 
\begin{align}
m_h^2 = -\lambda_{\Phi\varphi}(\mu=v')v'^2. 
\end{align}

$\eta$ is the inert doublet and the mass of $\eta$ should be positive.  
In our model, $\eta$ mass is generated by the mixing between $\eta$ and $\varphi$. 
Consequently, the mixing should be positive at the symmetry breaking scale, 
\begin{eqnarray}
\lambda_{\eta\varphi}>0. 
\label{inert1}
\end{eqnarray}
And the quartic couplings satisfy the following inert conditions\cite{Barbieri:2006dq}, 
\begin{eqnarray}
\lambda_\Phi>0,\  \lambda_\eta>0,\  
\lambda_{\Phi\eta}+\lambda'_{\Phi\eta}-\mid\lambda''_{\Phi\eta}\mid>-2\sqrt{\lambda_\Phi\lambda_\eta}. 
\label{inert2}
\end{eqnarray}

%The majorana mass term is generated by the majorana yukawa term. 

\subsection{Neutrino mass matrix}
The neutrino mass matrix can be obtained at one-loop level as follows~\cite{Ma:2006km, Hehn:2012kz}:
\begin{eqnarray}
({\cal M}_\nu)_{ab}=
\frac{(y_{\eta})_{ak}(y_{\eta})_{bk}M_k}{(4\pi)^2}
\left[\frac{m^2_R}{m^2_R-M^2_k}\ln\frac{m^2_R}{M^2_k}
%%%
-\frac{m^2_I}{m^2_I-M^2_k}\ln\frac{m^2_I}{M^2_k}
\right],
\end{eqnarray}
where $M_i\equiv (y_N)_iv'/\sqrt2$.
In this form, observed neutrino mass differences and their mixings %$U_{MNS}$
are obtained~\cite{Hehn:2012kz}, when the mixing matrix of the charged-lepton is diagonal basis. 
$Y_\eta$ can generally be written as
\begin{eqnarray}
Y_\eta=U^*_{MNS}
\left(
\begin{array}{ccc}
m_1^{\frac{1}{2}} & 0 & 0 \\
0 & m_2^{\frac{1}{2}} & 0 \\
0 & 0 & m_3^{\frac{1}{2}} \\
\end{array} 
\right)
O R^{\frac{1}{2}}, 
\label{yukawa}
\end{eqnarray}
where $U_{MNS}$ is the MNS matrix, $m_i$'s are neutrino masses, $O$ is an complex orthogonal matrix 
and $R$ is the following diagonal matrix, 
\begin{eqnarray}
R_{ii}=M_i\left(\frac{m_R^2}{m_R^2-M_i^2}\ln\frac{m_R^2}{M_i^2}-\frac{m_I^2}{m_I^2-M_i^2}\ln\frac{m_I^2}{M_i^2}\right).
\end{eqnarray}
We use this formula. We assume the lightest neutrino mass is zero and the neutrino mass spectrum is normal hierarchy. 
In this case, the complex orthogonal matrix $O$ can be written as 
\begin{eqnarray}
O=
\left(
\begin{array}{ccc}
0 & 0 & 1 \\
\cos\alpha & \sin\alpha & 0 \\
-\sin\alpha & \cos\alpha & 0 \\
\end{array} 
\right),
\label{co}
\end{eqnarray}
where $\alpha$ is complex parameter. 

\section{Numerical results}
We numerically solve the RGEs and find parameters that satisfy the inert conditions, 
Eq. (\ref{inert1}), (\ref{inert2}).

We use the following parameters at the Planck scale, 
\begin{eqnarray}
\lambda_\Phi=0.01, \  \lambda_\eta=0.09 ,\  \lambda_\varphi=0.011 ,\  \lambda_{\Phi\eta}''=10^{-9}, \ 
g_{B-L}=0.17 ,\  y_m=0.2,\ \alpha=0. 
\end{eqnarray}

%%%%%%%%%%%%%%%%%%%
\begin{figure}[cbt]
\begin{center}
%\unitlength=1mm
%\hspace{-2cm}
%\begin{picture}(40,100)
%\includegraphics[width=6cm]{neutrino.eps}
\includegraphics[scale=0.7]{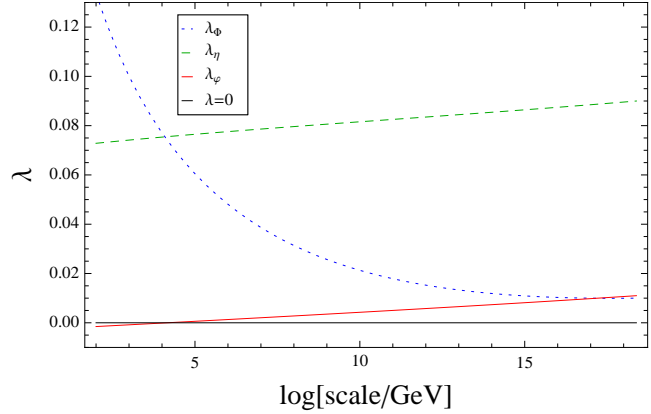}
%\qquad
% \includegraphics[scale=1]{neutrino2.eps}
%   \end{picture}
   \caption{Running for quartic couplings. Black solid line is $\lambda=0$ axis. }
   \label{rge1}
\end{center}
\end{figure}
%%%%%%%%%%%%%%%%%%%
%%%%%%%%%%%%%%%%%%%
\begin{figure}[cbt]
\begin{center}
%\unitlength=1mm
%\hspace{-2cm}
%\begin{picture}(40,100)
%\includegraphics[width=6cm]{neutrino.eps}
\includegraphics[scale=0.7]{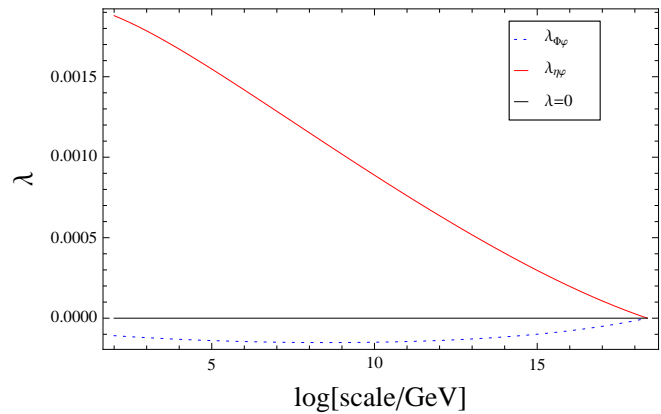}
%\qquad
% \includegraphics[scale=1]{neutrino2.eps}
%   \end{picture}
   \caption{Running for mixings between B$-$L Higgs and doublets. }
   \label{rge2}
\end{center}
\end{figure}
%%%%%%%%%%%%%%%%%%%
The Fig.~\ref{rge1} is the running for quartic couplings.
In this figure, $\lambda_\varphi$ becomes negative and satisfies Coleman-Weinberg condition (Eq.~(\ref{CWcon})) 
at $v'=$10.9 TeV. At that scale, other couplings satisfy inert conditions. 
The Fig.~\ref{rge2} is the running for mixings between B$-$L Higgs and doublets. 
In this case, Z' mass is 3.7 TeV. 
The experimental search for the Z' boson at LHC gives the limit on Z' boson mass, $m_{Z'}\geq 3$ TeV
\cite{Aad:2014cka, CMS:2013qca}.

\section{Conclusions}
We have investigated a classically conformal radiative seesaw model with gauged B$-$L symmetry, in which we have successfully obtained the 
B$-$L symmetry breaking through the Coleman-Weinberg mechanism. As a result, Majorana mass term is generated and
EW symmetry breaking occurs. We have also shown some allowed parameters to satisfy several constraints such as 
inert conditions, Coleman-Weinberg condition,  
the current bound on the $Z'$ mass at LHC, and so on as well as the neutrino oscillations experiments.

\begin{acknowledgments}
This work was supported by the Korea Neutrino Research Center which is established by the National Research Foundation of Korea(NRF) grant funded by the Korea government(MSIP) (No. 2009-0083526).
\end{acknowledgments}

%%%%%%%%%%%%%%%%%%%%%%%%%%%%%%%%%%

\bigskip % extra skip inserted

\end{document}